\pdfoutput=1 

\documentclass{article}
\usepackage{spconf,amsmath,graphicx}
\usepackage{lipsum}
\usepackage{url}
\usepackage{cleveref}
\usepackage{booktabs}

\makeatletter
\def\blfootnote{\gdef\@thefnmark{}\@footnotetext}
\makeatother

\title{DVDnet: A Fast Network for Deep Video Denoising}
\name{Matias Tassano$^{\star \dagger}$ \qquad Julie Delon$^{\star}$ \qquad Thomas Veit$^{\dagger}$}
\address{$^{\star}$ MAP5, Universit\'e Paris Descartes\\
	$^{\dagger}$ GoPro France}

\graphicspath{{./images/}}

\begin{document}
	%
	\maketitle

	\begin{abstract}
		In this paper, we propose a state-of-the-art video denoising algorithm based on a convolutional neural network architecture. Previous neural network based approaches to video denoising have been unsuccessful as their performance cannot compete with the performance of patch-based methods. However, our approach outperforms other patch-based competitors with significantly lower computing times. In contrast to other existing neural network denoisers, our algorithm exhibits several desirable properties such as a small memory footprint, and the ability to handle a wide range of noise levels with a single network model. The combination between its denoising performance and lower computational load makes this algorithm attractive for practical denoising applications. We compare our method with different state-of-art algorithms, both visually and with respect to objective quality metrics. The experiments show that our algorithm compares favorably to other state-of-art methods. Video examples, code and models are publicly available at \url{https://github.com/m-tassano/dvdnet}.
	\end{abstract}
	\begin{keywords}
		video denoising, CNN, residual learning, neural networks, image restoration\blfootnote{© 2019 IEEE.  Personal use of this material is permitted.  Permission from IEEE must be obtained for all other uses, in any current or future media, including reprinting/republishing this material for advertising or promotional purposes, creating new collective works, for resale or redistribution to servers or lists, or reuse of any copyrighted component of this work in other works.}
	\end{keywords}

	\section{Introduction}
	\label{sec:intro}
	
	We introduce a network for Deep Video Denoising: DVDnet. The algorithm compares favorably to other state-of-the-art methods, while it features fast running times. The outputs of our algorithm present remarkable temporal coherence, very low flickering, strong noise reduction, and accurate detail preservation.

	\subsection{Image Denoising}
	\label{sec:image-denoising}
	
	Compared to image denoising, video denoising appears as a largely underexplored domain. Recently, new image denoising methods based on deep learning techniques have drawn considerable attention due to their outstanding performance. Schmidt and Roth proposed in~\cite{Schmidt2014a} the cascade of shrinkage fields method that unifies the random field-based model and half-quadratic optimization into a single learning framework. Based on this method, Chen and Pock proposed in~\cite{Chen2017} a trainable nonlinear reaction diffusion model. This model can be expressed as a feed-forward deep network by concatenating a fixed number of gradient descent inference steps. Methods such as these two attain denoising performances comparable to those of well-known algorithms such as BM3D~\cite{Dabov2007a} or non-local Bayes (NLB~\cite{Lebrun2013c}). However, their performance is restricted to specific forms of prior. Additionally, many hand-tuned parameters are involved in the training process. In~\cite{Burger2012}, a multi-layer perceptron was successfully applied for image denoising. Nevertheless, a significant drawback of all these algorithms is that a specific model must be trained for each noise level.
	
	Another popular approach involves the use of convolutional neural networks (CNN), e.g.\ RBDN~\cite{Santhanam2016}, DnCNN~\cite{Zhang2017}, and FFDNet~\cite{Zhang2017a}. Their performance compares favorably to other state-of-the-art image denoising algorithms, both quantitatively and visually. These methods are composed of a succession of convolutional layers with nonlinear activation functions in between them. This type of architecture has been applied to the problem of joint denoising and demosaicing of RGB and raw images by Gharbi et al.\ in~\cite{Gharbi2016}. Contrary to other deep learning denoising methods, one of the remarkable features that these CNN-based methods present is the ability to denoise several levels of noise with only one trained model. Proposed by Zhang et al.\ in~\cite{Zhang2017}, DnCNN is an end-to-end trainable deep CNN for image denoising. This method is able to denoise different noise levels (e.g.\ with standard deviation $ \sigma \in [0, 55] $) with only one trained model. One of its main features is that it implements residual learning~\cite{He2016}, i.e.\ it estimates the noise existent in the input image rather than the denoised image. In a following paper~\cite{Zhang2017a}, Zhang et al.\ proposed FFDNet, which builds upon the work done for DnCNN\@.

	\subsection{Video Denoising}
	\label{sec:video-denoising}
	
	As for video denoising, the method proposed by Chen et al.\ in~\cite{chen2016deep} is one of the few to approach this problem with neural networks---recurrent neural networks in their case. However, their algorithm only works on grayscale images and it does not achieve satisfactory results, probably due to the difficulties associated with training recurring neural networks~\cite{pascanu2013difficulty}. Vogels et al.\ proposed in~\cite{vogels2018denoising} an architecture based on kernel-predicting neural networks able to denoise Monte Carlo rendered sequences. The state-of-the-art in video denoising is mostly defined by patch-based methods. Kokaram et al.\ proposed in \cite{kokaram1993motion} a 3D Wiener filtering scheme. We note in particular an extension of the popular BM3D to video denoising, V-BM4D~\cite{Maggioni2012}, and Video non-local Bayes (VNLB~\cite{Arias2018}). Nowadays, VNLB is the best video denoising algorithm in terms of quality of results, as it outperforms V-BM4D by a large margin. Nonetheless, its long running times render the method impractical---it could take several minutes to denoise a single frame. The performance of our method compares favorably to that of VNLB for moderate to large values of noise, while it features significantly faster inference times.
	
	\section{Our Method}
	\label{sec:method}
	\begin{figure*}[htb] 
		\begin{minipage}[b]{1.0\linewidth}
			\centering
			\centerline{\includegraphics[width=0.85\linewidth]{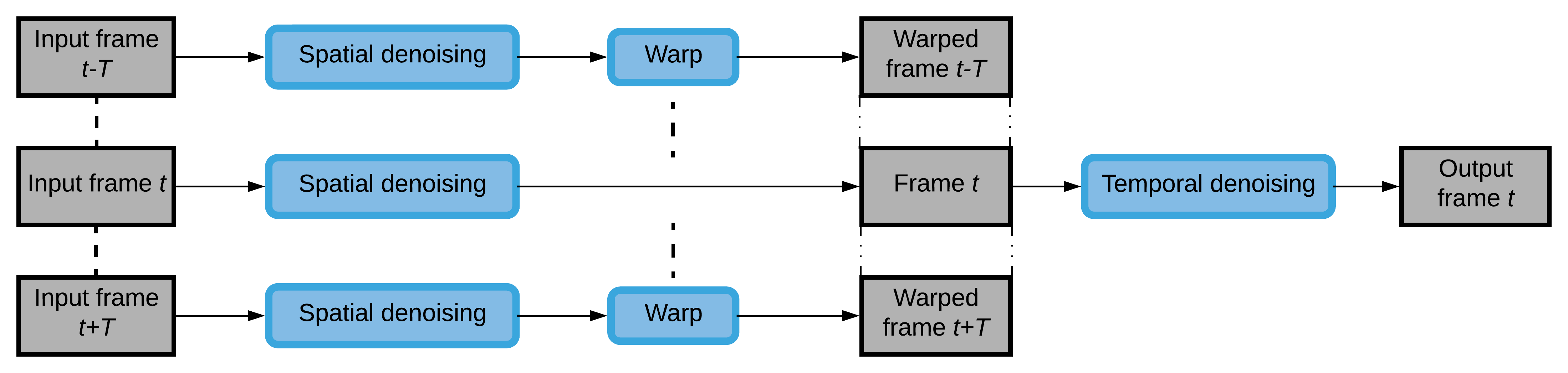}}
			%
		\end{minipage}
		\caption{Simplified architecture of our method.}
		\label{fig:architecture}
	\end{figure*}
	Methods based on neural networks are nowadays state-of-the-art in image denoising. However, state-of-the-art in video denoising still consists of patch-based methods. Generally speaking, most previous approaches based on deep learning have failed to employ the temporal information existent in image sequences effectively. Temporal coherence and the lack of flickering are vital aspects in the perceived quality of a video. Most state-of-the-art algorithms in video denoising are extensions of their image denoising counterparts. Such is the case, for example, of V-BM4D and BM3D, or VNLB and NLB. There are mainly two factors in these video denoising approaches which enforce temporal coherence in the results, namely the extension of search regions from spatial neighborhoods to volumetric neighborhoods, and the use of motion estimation. In other words, the former implies that when denoising a given pixel (or patch), the algorithm is going to look for similar pixels (patches) not only in the same frame, but also in adjacent frames of the sequence. Secondly, the use of motion estimation and/or compensation has been shown to help improving video denoising performance~\cite{Buades2016a,Arias2018,Maggioni2012}. We thus incorporated these two elements into our algorithm, as well as different aspects of other relevant CNN-based denoising architectures~\cite{Zhang2017a,Gharbi2016,vogels2018denoising}. Thanks to all these characteristics, our algorithm improves the state-of-the-art results, while featuring fast inference times.
	
	\Cref{fig:architecture} displays a simplified diagram of the architecture of our method. When denoising a given frame, its $ 2T $ neighboring frames are also taken as inputs. The denoising process of our algorithm can be split in two stages. Firstly, the $ 2T+1 $ frames are individually denoised with a spatial denoiser. Although each individual frame output at this stage features relatively good image quality, they present evident flickering when considered as a sequence. In the second stage of the algorithm, the $ 2T $ denoised temporal neighbors are registered with respect to the central frame. We use optical flow for this purpose. Splitting denoising in two stages allows for an individual pre-processing of each frame. On top of this, motion compensation is performed on pre-denoised images, which facilitates the task. Finally, the $ 2T+1 $ aligned frames are concatenated and input into the temporal denoising block. Using temporal neighbors when denoising each frame helps to reduce flickering as the residual error in each frame will be correlated. We also add a noise map as input to the spatial and temporal denoisers. The inclusion of the noise map as input allows the processing of spatially varying noise~\cite{ipol.2019.231}. Contrary to other denoising algorithms, our denoiser takes no other parameters as inputs apart from the image sequence and the estimation of the input noise.
	
	Observe that experiments presented in this paper focus on the case of additive white Gaussian noise (AWGN). Nevertheless, this algorithm can be straightforwardly extended to other types of noise, e.g.\ spatially varying noise (e.g.\ Poissonian). Let $ \mathbf{I} $ be a noiseless image, while $\tilde{\mathbf{I}}$ is its noisy version corrupted by a realization of zero-mean white Gaussian noise $ \mathbf{N} $ of standard deviation $ \sigma $, then
	\begin{equation}\label{eq:noisemod}
	\tilde{\mathbf{I}}=\mathbf{I}+\mathbf{N} \text{ .}
	\end{equation}

	\subsection{Spatial and Temporal Denoising Blocks}
	\label{sec:spatial-temporal-blocks}	
	
\begin{figure}[h]	
		\begin{minipage}[b]{1\linewidth}
			\centering
			\centerline{\includegraphics[width=0.9\linewidth]{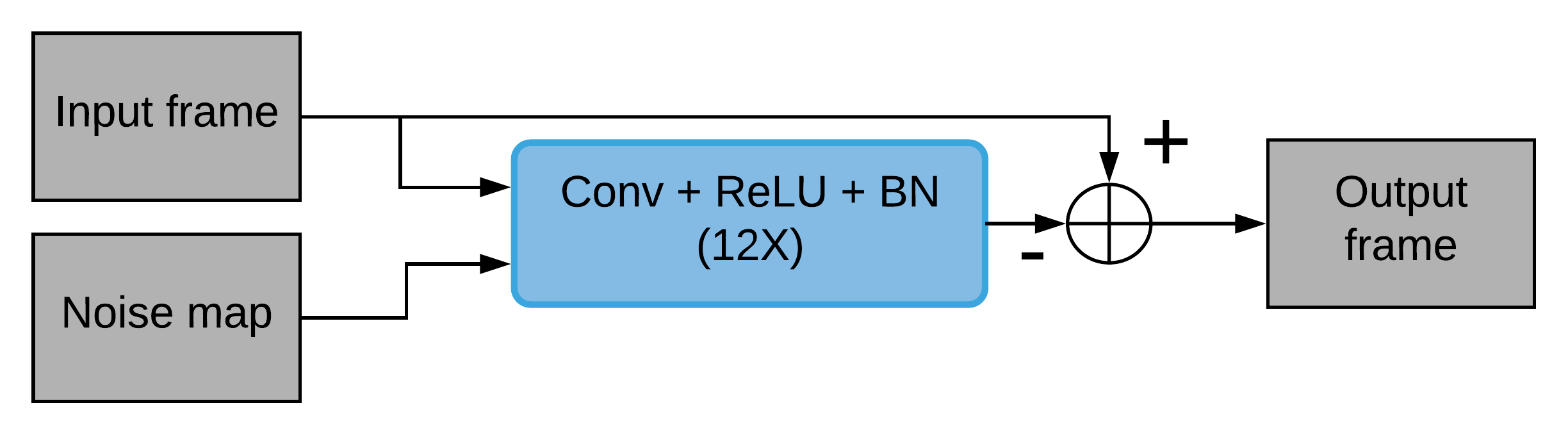}}
		\end{minipage}
		%
		\begin{minipage}[b]{1\linewidth}
			\centering
			\centerline{\includegraphics[width=0.9\linewidth]{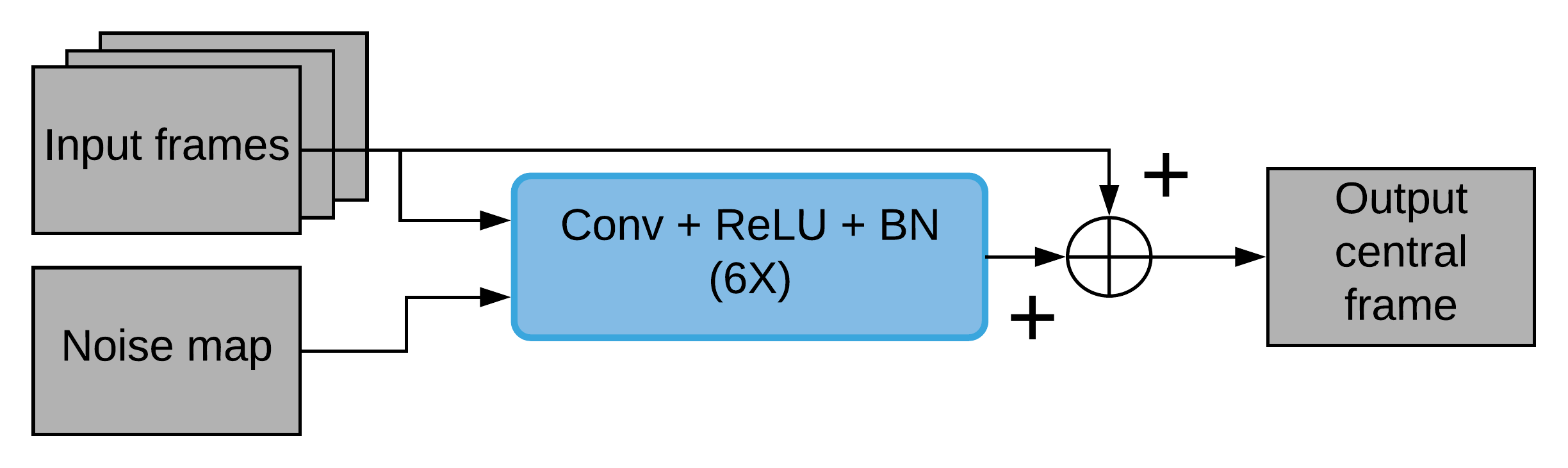}}
		\end{minipage}
		\caption{Simplified architecture of the spatial (top) and temporal (bottom) denoising blocks.}
		\label{fig:spatial-temporal-blocks}
	\end{figure}

	The design characteristics of the spatial and temporal blocks make a good compromise between performance and fast running times. Both blocks are implemented as standard feed-forward networks, as shown in \cref{fig:spatial-temporal-blocks}. The architecture of the spatial denoiser is inspired by the architectures in~\cite{Zhang2017a,Gharbi2016}, while the temporal denoiser also borrows some elements from~\cite{vogels2018denoising}.
	
 	The spatial and temporal denoising blocks are composed of $ D_{spa} = 12 $, and $ D_{temp} = 6 $ convolutional layers, respectively. The number of feature maps is set to $ W = 96 $. The outputs of the convolutional layers are followed by point-wise \textit{ReLU}~\cite{Krizhevsky2012} activation functions $ ReLU(\cdot) = \max (\cdot, 0) $. At training time, batch normalization layers (\textit{BN}~\cite{Ioffe2015}) are placed between the convolutional and \textit{ReLU} layers. At evaluation time, the batch normalization layers are removed, and replaced by an affine layer that applies the learned normalization. The spatial size of the convolutional kernels is $ 3 \times 3 $, and the stride is set to $ 1 $. In both blocks, the inputs are first downscaled to a quarter resolution. The main advantage of performing the denoising in a lower resolution is the large reduction in running times and memory requirements, without sacrificing denoising performance~\cite{Zhang2017a,ipol.2019.231}. The upscaling back to full resolution is performed with the technique described in~\cite{Shi2016}. Both blocks feature residual connections~\cite{He2016}, which have been observed to ease the training process~\cite{ipol.2019.231}.
	%
%
%
	\section{Training Details}
	\label{sec:training-details}
	
	The spatial and temporal denoising parts are trained separately, with the spatial denoiser trained first as its outputs are used to train the temporal denoiser. Both blocks are trained using crops of images, or patches. The size of the patches should be larger than the receptive field of the networks. In the case of the spatial denoiser, the training dataset is composed of pairs of input-output patches $ \left \{  \left( ( \mathbf{\tilde{I}}^j, \mathbf{M}^j ), \, \mathbf{{I}}^j \right )\right \}_{j=0}^{m_s} $ which are generated by adding AWGN with standard deviation $ \sigma \in [0, \, 55] $ to the clean patches $ \mathbf{{I}}^j $ and building the corresponding noise map $ \mathbf{M}^j $ (which is in this case constant with all its elements equal to $ \sigma $). A total of $ m_s = 1024000 $ patches are extracted from the Waterloo Exploration Database \cite{ma2017waterloo}. The patch size is $ 50 \times 50 $. Patches are randomly cropped from randomly sampled images of the training dataset. Residual learning is used, which implies that if the network outputs an estimation of the input noise $ \mathcal{F}_{spa} ( \,\mathbf{\tilde{I}} ; \theta_{spa}\,) = \mathbf{\hat{N}}
	 $, then the denoised image is computed by subtracting the output noise to the noisy input
	\begin{equation}\label{eq:res-den-img}
	\mathbf{ \hat{I}}( \,\mathbf{\tilde{I}} ; \theta_{spa}\,) = \mathbf{\tilde{I}} - \mathcal{F}_{spa} ( \,\mathbf{\tilde{I}} ; \theta_{spa}\,) \, .
	\end{equation}
	The loss function of the spatial denoiser writes
	\begin{equation}\label{eq:spa-loss}
	\mathcal{L}_{spa} (\theta_{spa})= \frac{1}{2{m_s}} \sum_{j=1}^{m_s} \left \| \mathbf{ \hat{I}}^j ( \,\mathbf{\tilde{I}}^j ;  \theta_{spa}\,) - \mathbf{I}^j \right \|^2 \, ,
	\end{equation}
	where $ \theta_{spa} $ is the collection of all learnable parameters.
	
	As for the temporal denoiser, the training dataset consists of input-output pairs $$ P_t^j = \left \{  \left( ( \,(^w\mathbf{\hat{I}}_{t-T}^j,\,\dots,\,\mathbf{\hat{I}}_{t}^j,\,\dots,\,^w\mathbf{\hat{I}}_{t+T}^j), \mathbf{M}^j\, ), \, \mathbf{{I}}_t^j \right )\right \}_{j=0}^{m_t} \, ,$$ where $ (^w\mathbf{\hat{I}}_{t-T}^j,\,\dots,\,\mathbf{\hat{I}}_{t}^j,\,\dots,\,^w\mathbf{\hat{I}}_{t+T}^j) $ is a collection of $ 2T+1 $ spatial patches cropped at the same location in contiguous frames. These are generated by adding AWGN of $ \sigma \in [0, \, 55] $ to clean patches of a given sequence, and denoising them using the spatial denoiser. Then, the $ 2T $ patches contiguous to the central reference patch $ \mathbf{{I}}_t^j $ are motion-compensated with respect to the latter, i.e.\ $ ^w{\mathbf{\hat{I}}_{l}^j} = \texttt{compensate}(\mathbf{\hat{I}}_{l}^j, \, \mathbf{\hat{I}}_{t}^j) $. To compensate frames, we use the DeepFlow algorithm~\cite{weinzaepfel:hal-00873592} for the estimation of the optical flow between frames. The noise map $ \mathbf{M}^j $ is the same as the one used in the spatial denoising stage. A total of $ m_t = 450000 $ training samples are extracted from the training set of the DAVIS database~\cite{KhoRohrSch_ACCV2018}. The spatial size of the patches is $ 44 \times 44 $, while the temporal size is $ 2T+1 = 5 $. The loss function for the temporal denoiser is
	\begin{equation}\label{eq:temp-loss}
	\mathcal{L}_{temp} (\theta_{temp})= \frac{1}{2{m_t}} \sum_{j=1}^{m_t} \left \| \mathbf{\hat{I}}_{temp, \,t}^j -  \mathbf{{I}}_{t}^j \right \|^2 \, ,
	\end{equation}
	where $ \mathbf{\hat{I}}_{temp,\,t}^j = \mathcal{F}_{temp} (P_t^j;\theta_{temp}) $.
	
	In both cases, the ADAM algorithm \cite{Kingma2015} is applied to minimize the loss function, with all its hyper-parameters set to their default values. The number of epochs is set to $ 80 $, and the mini-batch size is $ 128 $. The scheduling of the learning rate is also common to both cases. It starts at $ 1\mathrm{e}{-3} $ for the first $ 50 $ epochs, then changes to $ 1\mathrm{e}{-4} $ for the following $ 10 $ epochs, and finally switches to $ 1\mathrm{e}{-6} $ for the remaining of the training. Data is augmented five times by introducing rescaling by different scale factors and random flips. During the first $ 60 $ epochs, the orthogonalization of the convolutional kernels is applied as a means of regularization. It has been observed that initializing the training with orthogonalization may be beneficial to performance~\cite{Zhang2017a,ipol.2019.231}.

	\section{Results}
	\label{sec:results}
\begin{figure*}[htb] 
	\centering
	\begin{minipage}[b]{0.185\linewidth}
	\centering
	\centerline{\includegraphics[width=1\textwidth]{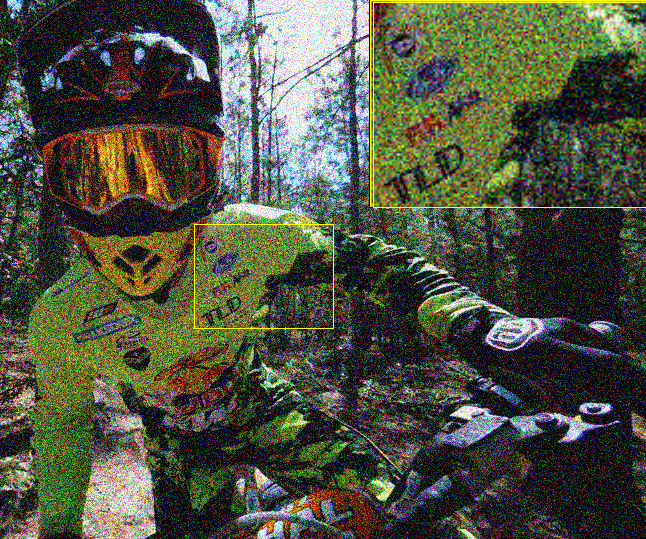}}
	\centerline{(b) Noisy $ \sigma = 50 $}\medskip
	\end{minipage}
	\begin{minipage}[b]{0.185\linewidth}
		\centering
		\centerline{\includegraphics[width=1\textwidth]{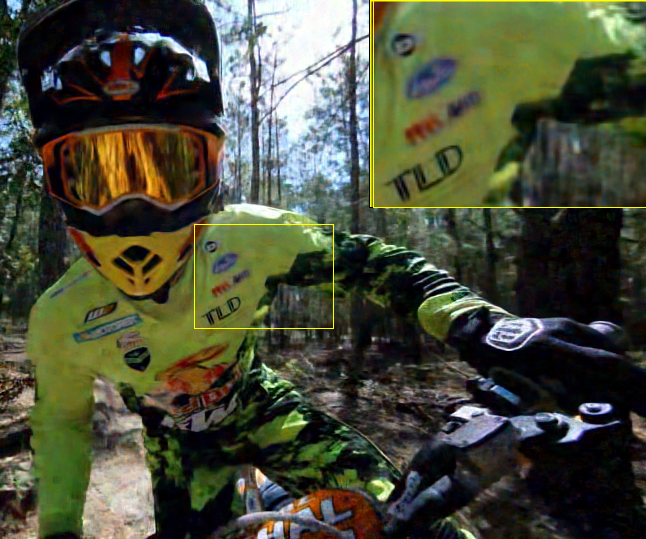}}
		\centerline{(c) V-BM4D}\medskip
	\end{minipage}
	\begin{minipage}[b]{0.185\linewidth}
		\centering
		\centerline{\includegraphics[width=1\textwidth]{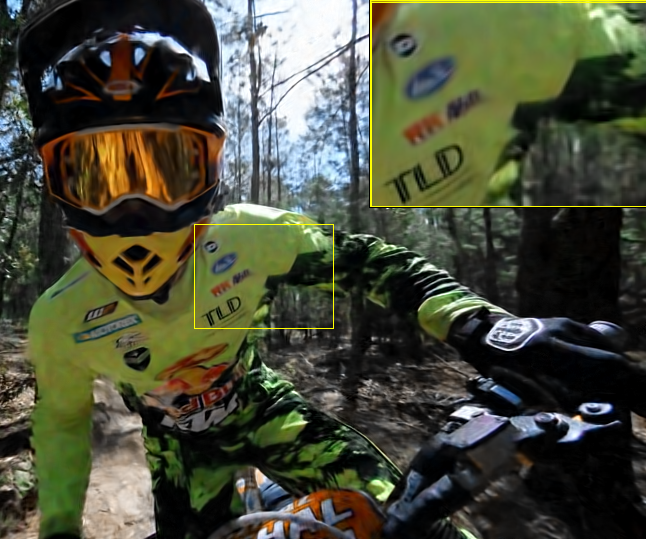}}
		\centerline{(d) VNLB}\medskip
	\end{minipage}
	\begin{minipage}[b]{0.185\linewidth}
		\centering
		\centerline{\includegraphics[width=1\textwidth]{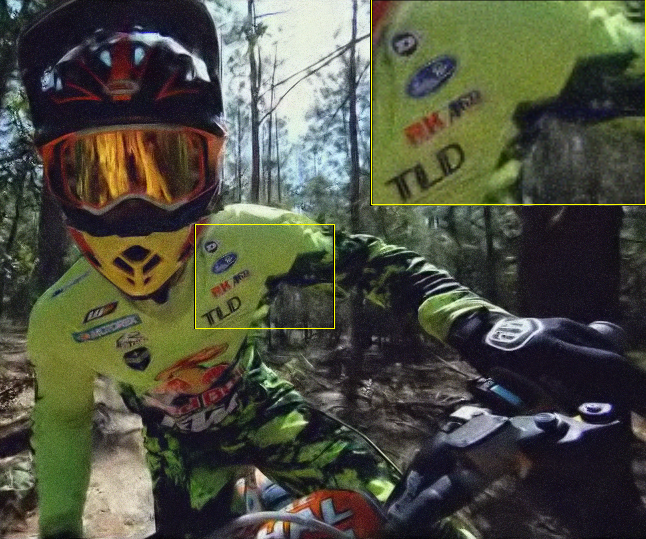}}
		\centerline{(e) Neat Video}\medskip
	\end{minipage}
	\begin{minipage}[b]{0.185\linewidth}
		\centering
		\centerline{\includegraphics[width=1\textwidth]{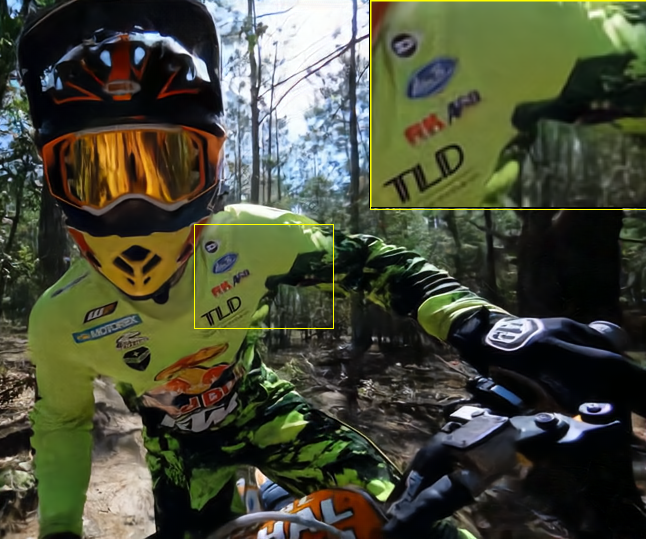}}
		\centerline{(e) DVDnet (ours)}\medskip
	\end{minipage}
	\caption{\textit{Comparison of results}. Left to right: noisy frame ($ \text{PSNR}_{seq} = 14.15dB $), output by V-BM4D ($ \text{PSNR}_{seq} = 24.91dB $), output by VNLB ($ \text{PSNR}_{seq} = 26.34dB $), output by Neat Video ($ \text{PSNR}_{seq} = 23.11dB $), output by DVDnet ($ \text{PSNR}_{seq} = 26.62dB $). Note the clarity of the denoised text, and the lack of low-frequency residual noise and chroma noise for DVDnet. Best viewed in digital format.}
	\label{fig:results}
\end{figure*}
	Two different testsets were used for benchmarking our method: the DAVIS-test testset, and Set8, which is composed of $ 4 $ color sequences from the \textit{Derf’s Test Media collection}\footnote{\url{https://media.xiph.org/video/derf}} and $ 4 $ color sequences captured with a GoPro camera. The DAVIS set contains $ 30 $ color sequences of resolution $ 854 \times 480 $. The sequences of Set8 have been downscaled to a resolution of $ 960 \times 540 $. In all cases, sequences were limited to a maximum of $ 85 $ frames. We used the DeepFlow algorithm to compute flow maps for DVDnet and VNLB. We also compare our method to a commercial blind denoising software, Neat Video (NV \cite{neatvideo19}).
		
	In general, DVDnet outputs sequences which feature remarkable temporal coherence. Flickering rendered by our method is notably small, especially in flat areas, where patch-based algorithms often leave behind low-frequency residual noise. An example can be observed in \cref{fig:results} (which is best viewed in digital format). Temporally decorrelated low-frequency noise in flat areas appears as particularly annoying in the eyes of the viewer. More video examples can be found in the website of the algorithm. The reader is encouraged to watch these examples to compare the visual quality of the results of our method.
	
	\Cref{tbl:results-set8,tbl:results-davis} show a comparison of $ PSNR $ on the Set8 and DAVIS dataset, respectively. It can be observed that for smaller values of noise, VNLB performs better. In effect, DVDnet tends to over denoise in some of these cases. However, for larger values of noise DVDnet surpasses VNLB.
	\begin{table}[!htbp]
		\centering	
		\caption{\label{tbl:results-set8}Comparison of $ PSNR $ on the Set8 testset.}		
			\begin{tabular}[b]{@{}l c c c c@{}l}
				\toprule[0.8pt]
				  & DVDnet & VNLB & V-BM4D & NV \\ 
				\midrule[0.8pt]
				$ \sigma = 10 $ & 36.08 & \textbf{37.26} & 36.05 & 35.67 \\ 
				$ \sigma = 20 $ & 33.49 & \textbf{33.72} & 32.19 & 31.69 \\ 
				$ \sigma = 30 $ & \textbf{31.79} & 31.74 & 30.00 & 28.84 \\ 
				$ \sigma = 40 $ & \textbf{30.55} & 30.39 & 28.48 & 26.36 \\ 
				$ \sigma = 50 $ & \textbf{29.56} & 29.24 & 27.33 & 25.46 \\ 
				\bottomrule[0.8pt]& 
			\end{tabular}
	\end{table}
	\begin{table}[!htbp]
		\centering
		\caption{\label{tbl:results-davis}Comparison of $ PSNR $ on the DAVIS testset.}	
			\begin{tabular}[b]{@{}l c c c}
				\toprule[0.8pt]
				  & DVDnet & VNLB & V-BM4D \\ 
				\midrule[0.8pt]
				
				$ \sigma = 10 $ & 38.13 & \textbf{38.85} & 37.58 \\ 
				$ \sigma = 20 $ & \textbf{35.70} & 35.68 & 33.88 \\ 
				$ \sigma = 30 $ & \textbf{34.08} & 33.73 & 31.65 \\ 
				$ \sigma = 40 $ & \textbf{32.86} & 32.32 & 30.05 \\ 
				$ \sigma = 50 $ & \textbf{31.85} & 31.13 & 28.80 \\ 

				\bottomrule[0.8pt]& 
			\end{tabular}
	\end{table}
	\subsection{Running times}
	\label{ssec:running-times}
	
	Our method achieves fast inference times, thanks to its design characteristics and simple architecture. DVDnet takes less than $ 8s $ to denoise a $ 960 \times 540 $ color frame, which is about $ 20 $ times faster than V-BM4D, and about $ 50 $ times faster than VNLB. Even running on CPU, DVDnet is about an order of magnitude faster than these methods. Of the $ 8s $ it takes to denoise a frame, $ 6s $ are spent on compensating motion of the temporal neighboring frames. \Cref{tbl:running-times} compares the running times of different state-of-the-art algorithms.
	\begin{table}[!htbp]
		\centering
		\caption{\textit{Comparison of running times.} Time to denoise a color frame of resolution $ 960 \times 540 $. Note: values displayed for VNLB do not include the time required to estimate motion.}
		\begin{minipage}[b]{0.8\linewidth}
			\centering
			\begin{tabular}{@{}l c c c c@{}}
				\toprule[1pt]
				\textbf{Method} & V-BM4D & VNLB & DVDnet & DVDnet \\
				&  &  & (CPU) & (GPU)\\
				\midrule[0.8pt]
				\textbf{Time (s)} & 156 & 420 & 19 & 8 \\
				\bottomrule[0.5pt]
			\end{tabular}
			
		\end{minipage}
		\label{tbl:running-times}
	\end{table}
	%
	%
	%
	
	\section{Conclusions}
	\label{sec:conclusions}
	
	In this paper, we presented DVDnet, a video denoising algorithm which improves the state-of-the-art. Denoising results of DVDnet feature remarkable temporal coherence, very low flickering, and excellent detail preservation. The algorithm achieves running times which are at least an order of magnitude faster than other state-of-the-art competitors. Although the results presented in this paper hold for Gaussian noise, our method could be extended to denoise other types of noise.
		
	\vfill\pagebreak
	\bibliographystyle{IEEEbib}
	\bibliography{refs}
	
\end{document}